\documentclass[preprint,12pt,3p]{elsarticle}

\usepackage{amssymb}

 \usepackage{lineno}

\journal{Elsevier}
\usepackage{color}
\usepackage{amssymb}
\usepackage{array}
\usepackage{booktabs}
\usepackage{multirow}
\usepackage{subfigure}
\usepackage{graphicx}
\usepackage{dcolumn}
\usepackage{bm}
\usepackage{mathtools}

\begin{document}

\begin{frontmatter}

\title{Growth evolution of self-affine thermally evaporated KBr thin films: A fractal assessment}

\author[add1]{R. Rai}
\author[add2]{R. P. Yadav}
\author[add1,add3]{Triloki}
\author[add1]{Nabeel Jammal}
\author[add1]{A. K. Singh}
\author[add1]{B. K.~Singh\corref{cor}}
\cortext[cor]{Corresponding author}
\ead{bksingh@bhu.ac.in}

\address[add1]{Department of Physics, Institute of Science, Banaras Hindu University, Varanasi-221005, India.}
\address[add2]{Department of physics, MNNIT, Allahabad-211004, India.} 
\address[add3]{Present address: Research Center for Nuclear Physics, Osaka University, Ibaraki, Osaka-567-0047, Japan.}

\begin{abstract}

 In this article, fractal concepts were used to explore the thermally evaporated potassium bromide thin films of different thicknesses 200, 300, and 500 nm respectively; grown on aluminium substrates at room temperature. The self-affine or self similar nature of growing surfaces was investigated by autocorrelation function and obtained results are compared with the morphological envelope method. Theoretical estimations revealed that the global surface parameters such as, interface width and lateral correlation length are monotonically decreased with increasing film thickness. Also, from height profile and A-F plots, it has been perceived that irregularity/ complexity of growing layers was significantly influenced by thickness. On the other hand, the fractal dimension and local roughness exponent, estimated by height-height correlation function, do not suggest such dependency.
 \end{abstract}


\end{frontmatter}


\section{Introduction}
\label{sec1}

From last few decades, alkali halides (A-H) thin films are extensively used to tailor the quantum detection efficiency of the extreme ultraviolet (EUV, 10 nm \textless $\lambda$ \textless 100 nm) and vacuum ultraviolet (VUV, 100 nm \textless $\lambda$ \textless 200 nm) sensitive photodetectors for the high energy and astrophysics experiments. Applications of A-H films are  also diversified over the many other sophisticated fields like; medical imaging~\cite{WeiZhao}, positron emission tomography~\cite{F.Garibaldi}, ultrafast electron microscope (UEM)~\cite{Sergei}, generation of highly instance free electron laser beams~\cite{J. Workman}, scintillation detectors~\cite{Daisuke}, etc.  In particular, at the shorter wavelength ranges ($\lambda_{cutoff}$ $<$160 nm), potassium bromide (KBr) thin films have been proved to be a very efficient photoconverter and employed in various field including its use as a protecting layers in the visible sensitive photon counting devices~\cite{Oswald,ASTr} and also in various  planetary exploration mission such as the UV spectroscope PHEBUS on Board of ESA/JAXAs BepiColombo Mercury exploration mission~\cite{PHEBUS}, SUMER on the Solar and Heliospheric Observatory (SOHO)~\cite{SUMER}, the Far Ultraviolet Spectroscopic Explorer (FUSE)~\cite {over}, the Cassini Ultraviolet Imaging Spectrograph XUV on the NOZOMI~\cite{UVIS}. It is an established fact that for the stable operation of large area UV devices, there will always be a requirement of homogeneous thin film coating, which  permits simple handling and withstand with moderate amount of exposure towards humidity and high energetic photons/ion bombardment. Therefore, the optimization of film fabrication process is prerequisite to ensure high response and reproducibility of the UV devices. In connection to this, the morphological characterization probes have been emerged as a powerful tools to examine the film quality in terms of surface coverage and homogeneity. 

A variety of deposition methods like, chemical vapour deposition, ion beam sputtering~\cite{IonBeam}, electron beam evaporation~\cite{Egun}, pulsed laser deposition~\cite{pulse_laser}, spray pyrolysis~\cite{spraypyrolysis} and  ionic liquid growth~\cite{IonicLiduid} can be utilized to synthesize alkali halide thin films.  The sample prepared by different deposition methods follow a very different growth mechanism, however, this article only considers the evolution in surface morphology originated from the non energetic thermal evaporation process. Due to a simple deposition procedure, the thermal evaporation technique is a widely employed coating process; frequently used for growing KBr films for their different applications. Despite of this, very less work available in literature which aimed a surface analysis of thermally grown films. Therefore, it seems reasonable to analyse the growth evolution quantitatively and systematically to map the surface morphology of these films.

Generally, the surface morphology of films is manifested by complex irregular microstructures, whose geometry cannot be quantified by traditional Euclidean approach. The non-equilibrium growth fronts exhibits a random self-affine behaviour in spatial domain as well as in temporal domain ~\cite{Barabasi}. In this context, the scaling and fractal geometry concept has been evolved as a successful tool to explore the structural complexity of thin films ~\cite{Mandelbrot}. Several researchers have been applied height-height correlation function, to describe the surface parameters like roughness exponent, interface width, lateral correlation length and fractal dimension of such fracture surfaces. For example, Yadav et. al.~\cite{Yadav_2012} calculated the fractal dimension and the roughness exponent for LiF thin films deposited by electron beam evaporation techniques with varying thicknesses. They found that the grain size, fractal dimension and roughness exponent are thickness dependent~\cite{Yadav_2012}. Similar observation reported by Maryam et. al.~\cite{Maryam Nasehnejad} for Au films, prepared by electrodeposition method i.e. fractal dimension and lateral correlation length is strongly influenced by thickness. A strong correlation between electrical resistivity and fractal dimension for Ag/Cu thin film was reported by Talu et al.\cite{Talu_2017}. Their study reported that fractal dimension and the electrical resistance decreased as the thin film thickness increases. Singh et al.~\cite{U.B.Singh} used fractal technique to study the insight mechanisms of Ag thin films and it's surface structuring. Gupta et. al.~\cite{Gupta} also used height-height correlation function to quantify the surface parameter of CdS thin film grown from chemical bath and show the anomalous scaling pattern with rapid roughening of film surfaces as a consequence of bulk diffusion instability. In their study of prylene films, grown on Si substrate Serkan Zorba et. al.~\cite{Zorba} found that the height-height correlation function saturates once nominal thickness is reached to its the critical value. As illustrated from literatures that extensive studies have been carried out to understand the surface dynamics of thin films deposited from different materials. In spite of fact, this is the first fractal analytical study of surface roughness of KBr films and it will provide a new passage for both characterization and direct prediction about the surface properties of thermally grown layers.

In this article, we present a detailed analysis aimed at elucidating the variation in the growth parameters of a thermally deposited KBr films of different thickness. We employed Atomic Force Microscope (AFM) technique to generate a relevant statistical data required for characterization of the growth dynamics in terms of roughness and lateral correlation length.

\begin{figure}[!ht]
 \begin{center}
\includegraphics[scale=0.75]{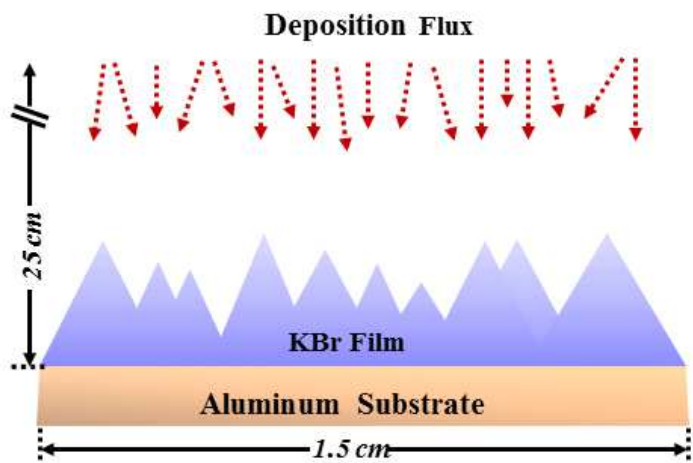} 
 \caption{ Schematic representation of growth evolution during evaporation process.}
     \label{TH}
    \end{center}
  \end{figure}

\section{Experimental Details}
 Thin films were fabricated in a controlled vacuum environment by evaporating high purity KBr powder (99.99\%, Alfa Aesar) from a electrically heated tantalum ($Ta$) boat. The ambient pressure of deposition chamber was reduced by a turbomolecular pumping system (speed: 510 L/s) that was baked with a XtraDry pump (Pfeiffer) . At the ultimate pressure $\sim 5.6\times 10^{-7}$ Torr, the composition of vacuum chamber was tracked by a residual gas analyzer between 1 to 90 amu mass ranges and only the fragmentation patterns of some commonly present molecules; $H_{2}^{+}$: 33.9\%, $N_{2}^{+}$: 40.5\%, $O_{2}^{+}$: 2.0\%, $H_{2}O^{+}$: 21.6\% were observed. At this stage, the deposition started  by applying an 80 Amp. current on a $Ta$  boat and collecting an evaporated flux on to an aluminium substrate which is kept at the 25 cm height from a KBr source. The controlled deposition has been performed by monitoring the evaporation rate ($\leq 3nm/s$) and film thickness using a $Quartz$ crystal based thickness/rate monitor. After the deposition of desired thickness, samples are transferred into a vacuum desiccator under the constant flow of dry $N_{2}$ gas and immediately transported for further characterization.

 AFM, NT-MDT solver-NEXT~\cite{AFM} in semi contact mode, was conceived to acquire submicron images of KBr film. To assure that the observed morphology is representative of entire film surface few randomly selected region are scanned and results are found to be consistence.  All the AFM measurements were carried out at room temperature (RH=45\%) over the $5\mu m\times 5\mu m$ with the resolution of 256 pixels$\times$ 256 pixels for each film. Data acquisition and off line analysis were carried out by WSXM~\cite{WSXM} and MountainMap$^{^{\textcircled{\tiny{R}}}}7$ (Digital Surf)~\cite{Mountains} image processing softwares.   Fig.~\ref{TH} displays, the schematic of KBr thin film surfaces after evaporation.

\section{Statistical Analysis}
AFM images have been quantitatively analysed by MountaiMap$^{^{\textcircled{\tiny{R}}}}7$ software. Apart from this, following mathematical algorithms are also used to present a detail description of digitized film surfaces. 

\subsection{Interface width} 
Experimentally, the surface height is measured over a discrete lattice and the height of the surface at point $(i,j)$ is denoted by $h(i,j)$. The root mean square (RMS) value of the deviation of the surface height  from its mean is known as interface width $w$ which is given by: 
                                                                
\begin{equation}
w = \bigg\{\frac{1}{N^{2}}\sum_{i=1}^N \sum_{j=1}^N[h(i,j)-\langle h(i,j)\rangle]^{2}\bigg\}^{\frac{1}{2}}
\end{equation}
act mode 
where $\langle h(i,j)\rangle$ corresponds to the mean value of the heights over a square surface and is denoted by 

\begin{equation}
\langle h(i,j)\rangle = \frac{1}{N^{2}}\sum_{i=1}^N \sum_{j=1}^N h(i,j)
\end{equation}

\subsection{Correlation properties and Fractal Dimension}

The interface width is a global parameter and it is only characterizes the shape of peaks and valleys of the surfaces but it does not give any information about correlation and self-affine (self-similar) properties. To describe the correlation properties, it is normally needed to redefine surface heights. For this purpose the mean height $h(i,j)$ is taken to be zero, so that $h(i,j)$ denotes the fluctuation of the surface height from the mean at the position $(i,j)$. The correlation property in AFM images along fast scan direction is described by autocorrelation function,  
                
\begin{equation}
A(r=md) = \frac{1}{N(N-m)w^{2}}\sum_{i=1}^N \sum_{j=1}^{N-m} h(i+m,j)h(i,j)
\end{equation}

For thin film surfaces, $A(r)$ is often found to have exponentially decreasing behaviour. The value $\xi$,  for which $A(\xi)$  drops $1/e$ of its original value is known as lateral correlation length. Therefore,

\begin{equation}
A(\xi) \cong \frac{1}{e}
\end{equation}

Height-height correlation function $H(r)$ is another function which is applied to describe the correlation property of surface. $H(r)$ in fast scan direction is given by 

\begin{equation}
\begin{split}
H(r) & = {\frac{1}{N(N-m)}\sum_{i=1}^N \sum_{j=1}^{N-m}[h(i+m,j)-h(i,j)]^{2}}\\ 
    &  = 2w^{2}[1-A(r)]
\end{split}
\end{equation} 

Using Eq.(4) we have,

\begin{equation}
H(r)= 2w^{2}(1-1/e)
\end{equation}

For self-affine surfaces, $H(r)$ behaves as 
  
\begin{equation}
  H (r) =\begin{cases}
      2w^{2} ~ \text{for r $>>$ $\xi$}\\
      r^{2\alpha}~ \text{for r $<<$ $\xi$}
       \end{cases}       
\end{equation}

where $\alpha$ is  known as roughness exponent, which characterizes the short range roughness of self-affine surface. The larger value of  $\alpha$ represents a smoother local surface profile.
The value of $\alpha$  is directly related to fractal dimension $(D_{f})$ as

\begin{equation}
D_{f} = d+1-\alpha
\end{equation} 

where $d$, is the dimension of sample. In our case it is equal to 2~\cite{Yadav_2015}.

\section{Results and Discussion}

\subsection{AFM analysis}

\begin{figure}[ht!]
 \begin{center}
\includegraphics[scale=0.63]{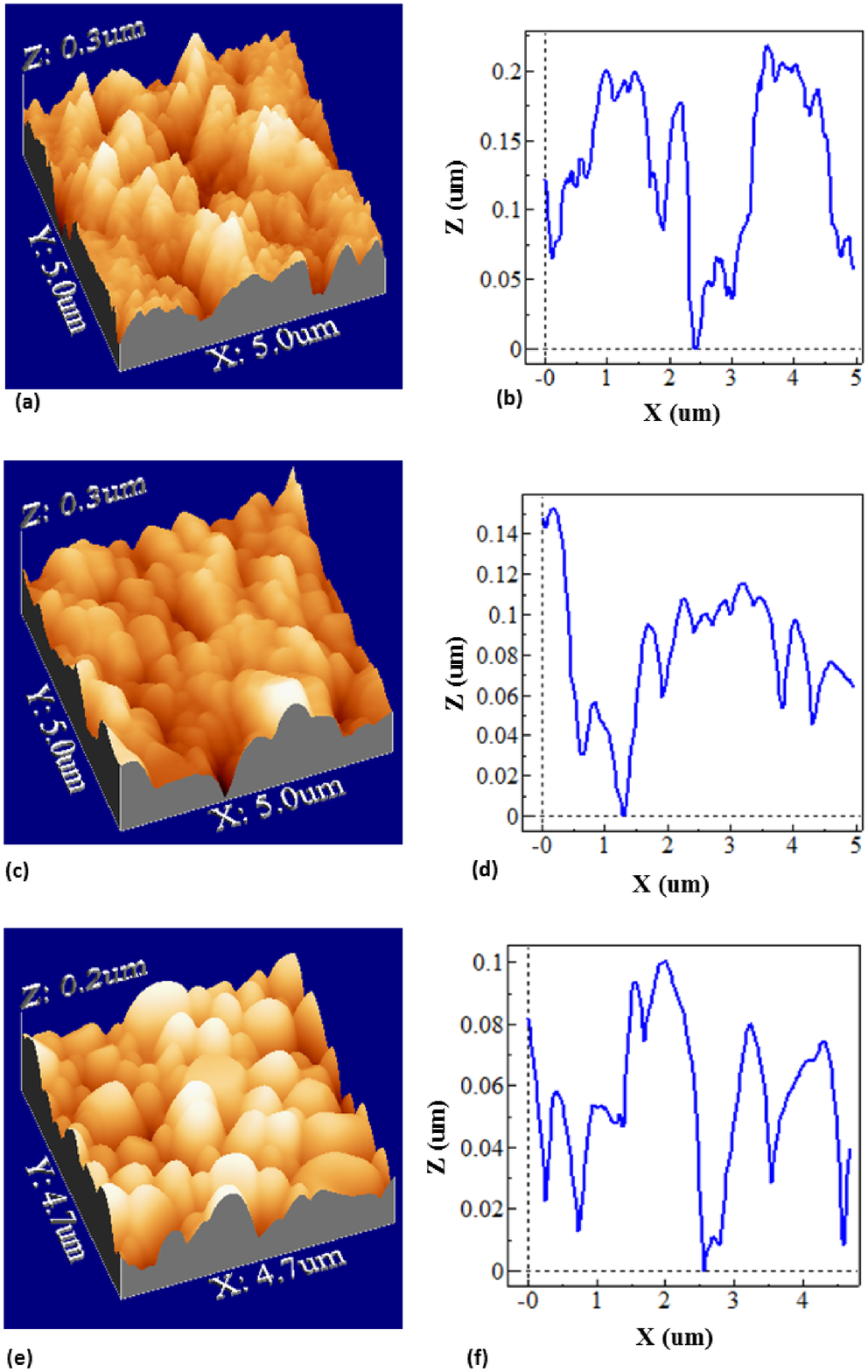} 
 \caption{ 3D AFM images (a), (c), (e) and height profiles (b), (d) (f) of 200, 300 and 500 nm thick KBr films across the $5\mu m \times 5\mu m$ scan area.}
     \label{XRD}
    \end{center}
  \end{figure} 

Fig. 2 illustrates, typical surface morphology of the KBr films with increasing thickness and corresponding height profile along the x-axis. It is evident from the images that the morphology of film's surfaces is mainly formed by a non uniform distribution of granular structure and with the elevation of the deposition time, $i.e.$ film thickness, these island are grow in both direction; vertically as well as horizontally.  The 200 nm film is constituted from a discontinuous distribution of sharp edge columns, in contrary for thicker films these column become denser. Since the microstructural growth of KBr films follow a Volmer-Weber growth mode and in this mode films initially grow by the nucleation of discrete islands of different crystallographic orientations, which is visible for a 200 nm thick film in Fig 2(a). With further deposition, the existing grains enlarge, and new islands are also nucleate, till than a continuous percolating interconnected channels of columns are formed, which is visible in the case of 300 nm and 500 nm films. Therefore with increasing film thickness the substrate surface coverage area and grain size increased as observed for 300 nm and 500 nm thick films. Similarly, the height profile of 200 nm film shows many bumps in the range of 20 to 25 nm, however for 300 nm and 500 nm films, widths of these bumps have been increased while height variation decreases. It may be predicted that the large scale variation in height parameter will be decreased with the increment in thickness.

 The polar representations for texture analysis of KBr films are illustrated in Fig. 3(a)-(b) that dominant growth direction appears on the film surface is more or less perpendicular to the substrate, which results in the columnar structure with the preferred crystallographic texture. This fact  is also supported by the cross sectional view of film's surfaces that the dominant growth direction appears on the film surface
is at $90^{o}$ from the substrate.
\begin{figure}[!ht]
 \begin{center}
\includegraphics[scale=0.37]{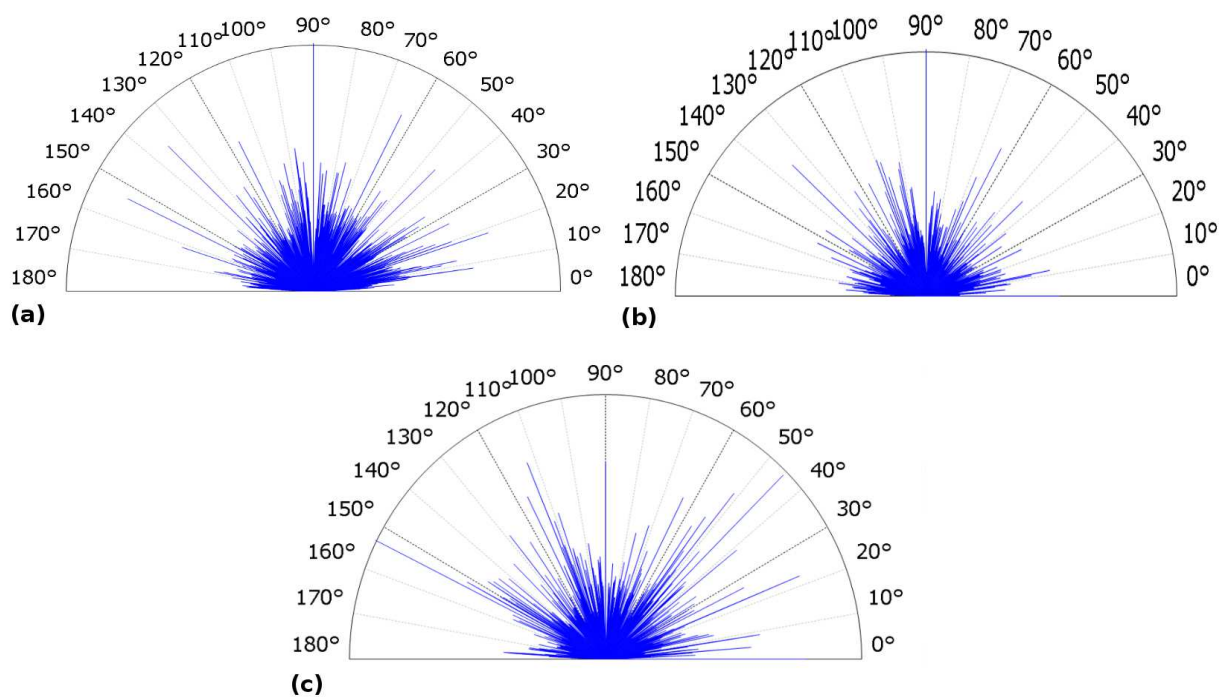} 
 \caption{The polar representation for texture analysis of (a) 200, (b) 300, and (c) 500 nm thick KBr films.}
     \label{XRD}
    \end{center}
  \end{figure}

\subsection{The depth histogram profile of rough surfaces}

\begin{figure}[!ht]
 \begin{centering}
\includegraphics[scale=0.72]{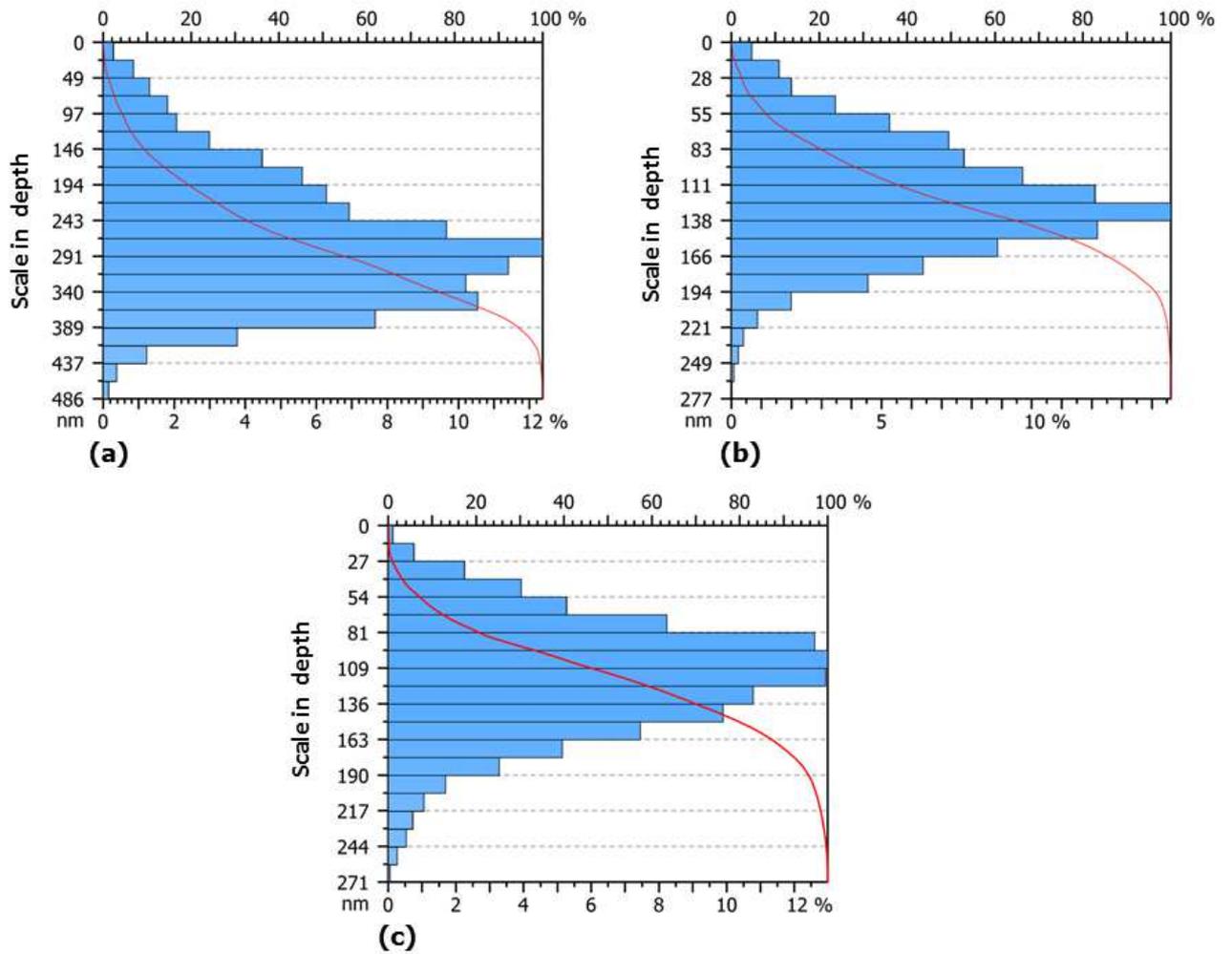} 
 \caption{The depth histogram for the KBr films: (a) 200 nm, (b) 300 nm and (c) 500 nm. The Abbott–Firestone curve is indicated by red
color.}
     \label{XRD}
    \end{centering}
  \end{figure} 

The depth histograms associated with $3D$ AFM images (Fig. 2(a), (c) and (d)) of KBr layers, deposited on aluminium substrates are shown in the Fig. 4(a)-(c). These depth histograms illustrate the density dispensation of the data points on the film surface. The vertical axis is scaled in depths and the horizontal axis in the percent of the total material population. The Abbott-Firestone curve is overlaid in red colour and curve characterized roughness, or bearing capacity of the material; by providing the statistical information of the distribution of material in the length range of profile. This curve is graphically described that for a specific depth, the percentage of the material traversed in relation to the covered area. Material ratio curve (Fig. 4, red plot), is the integral of the amplitude distribution function (ADF/Surface Histogram). It is a cumulative probability distribution and a measure of the material to air ratio expressed as a percentage at a particular depth below the highest peak in the surface. The curve starts from the highest peak, where material is 0\% and ends in the lowest valley, where  material is 100\% and give access to  both  size  and proportion   of   the   peaks   and   valleys observed on the material   surface. From A-F curve, a stable value of roughness height is evaluated and comes out to be about 285, 189, 138 nm for 200, 300 and 500 nm film respectively. This calculation excludes the highest peaks that will be worn out and the deepest valleys that will be filled in.

\subsection{Fractal analysis}
 Fractal analysis of the stereometric data was conducted based on algorithms discussed in section 3.1 \& 3.2, which consists of fractal scaling of the surface measured with an AFM. From the images, both surface roughening and coarsening process are evident that is occurred due to increase in film thickness. 
 
The interface width $w_{Th}$, which is a global parameter and measures the height fluctuation of the surface, is computed using the relation describe in section 3.1 for the thin films of thicknesses 200 nm, 300 nm and 500 nm, respectively. The computed values of $w_{Th}$ are  0.0694, 0.0331 and 0.0283 $\mu m$, respectively. From the $w_{Th}$ values, it is observed that as the thickness of films increases, the value of $\xi_{Th}$ monotonically decreases. This results is also supported by height profile and A-F plots of KBr films, in which it is expected that with increasing film thickness roughness will be reduced.  As such smoothness and maximum surface coverage are desirable to reduce the possibility of efficiency loss through the excessive scattering and less absorption. Therefore, it may improve the performance and sensitivity of UV devices. This observation is also supported by our earlier work, in which we found that with increasing film thickness, photoemission has been increased~\cite{Richa} But they do not provide any information about complexity of surfaces. These values are only sensitive to the peak and valley values of surface profile. The self-affine roughness is widely characterized by engaging it to a dynamic scaling form like autocorrelation function $A(r)$ and the height-height correlation function $H(r)$. 

Normalized autocorrelation function $A(r)$ described in section 3.2 is employed to confirm that the surface under investigation have self-affine or not. The exponentially decreasing behaviour of $H(r)$ verify that the surface under investigation have self-affine nature.  We plotted a graph between  $A(r)$  and $r$  as shown in Fig. 5. The value of $\xi_{Th}$ is computed for each thin film.  The computed values $\xi_{Th}$ of  are 0.6867, 0.3217 and 0.3094 $\mu m$, respectively. It is observed that as the thickness of the film increases the value of $\xi_{Th}$ is decreased. A crossover occurs at $r=\xi_{Th}$ for each thin film, which suggests that the distance at which the surface features are no longer correlated. Thus, $A(r)$ provides vertical as well as lateral information about the surfaces ~\cite{RPYadav, Yadav_2015, Yadav}.
 
Quantitative information of the surface morphology can be extracted from the height-height correlation function $H(r)$. The relation between  $A(r)$ and $H(r)$  is described by Eqn.(5). We computed $H(r)$ for each thin film surface. The two distinct regions are observed in $log~H(r)$
versus $log~r$  plot for each thin film surface as shown in Fig. 6. The linear portion of  $log~H(r)$ versus $log~r$ plot shows power-law $i.e.$ $H(r)\sim r^{2\alpha}$ for small $r$, while for larger $r$ quasi-periodic behaviour is observed.  The difference in behaviour arises at length scale beyond $r>>\xi_{Th}$.  Self-similar surface shows constant behaviour in this region $r>>\xi_{Th}$ and becomes quasi-periodic for oscillatory surface.  It is also important to mention that the surface will only reveal self-affine behaviour over certain range of length scales.  All $H(r)$ do not overlap to each other, which suggest that the growth is non stationary and the scaling relation $Z=\alpha/ \beta$  may break down ~\cite{Karabacak}.

The roughness exponent ($\alpha_{_{Th}}$) is determined from a fit to the linear part of the log-log plot of $H(r)$ versus $r$. The computed values of  $\alpha_{_{Th}}$ are 0.8022, 0.8869 and 0.8590 respectively which quantify the roughness changes with length scale and is indicative of surface texture. It also describes the surface fractality. The value of $\alpha_{_{Th}}$ measures the sharp local surface irregularities with length scale. For $\alpha_{_{Th}} = 1$, the surface looks smoother while, $\alpha_{_{Th}} < 1$ surface is looking rough. The fractal analysis is a suitable approach for characterization and understanding of surface morphology. The relation between $\alpha_{_{Th}}$ and $D_{f_{Th}}$ is described in section 3.2. We find that the fractal dimension is  2.1978, 2.1131,  and 2.1410 respectively which describes the total profile complexity and reveals the change of normalized profile length with observational scale.

\begin{figure}[!ht]
 \begin{center}
\includegraphics[scale=0.65]{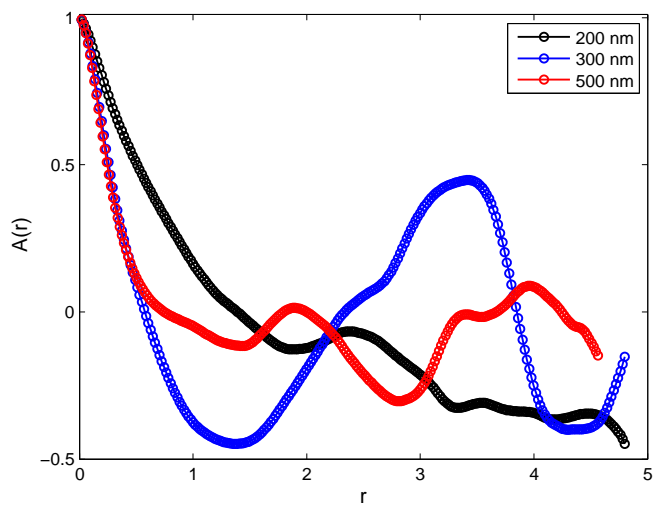}  
 \caption{Autocorrelation function A(r) versus r for 200 nm,  300 nm and 500 nm KBr films.}
     \label{XRD}
    \end{center}
  \end{figure}

\begin{figure}[!ht]
 \begin{center}
\includegraphics[scale=0.65]{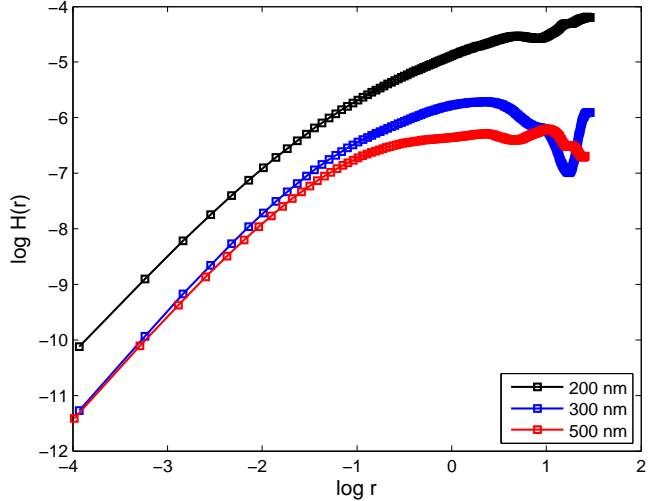} 
 \caption{log H(r) as a function of log r for each thickness of KBr films.}
     \label{XRD}
    \end{center}
  \end{figure}

\begin{figure}[!ht]
 \begin{centering}
\includegraphics[scale=0.80]{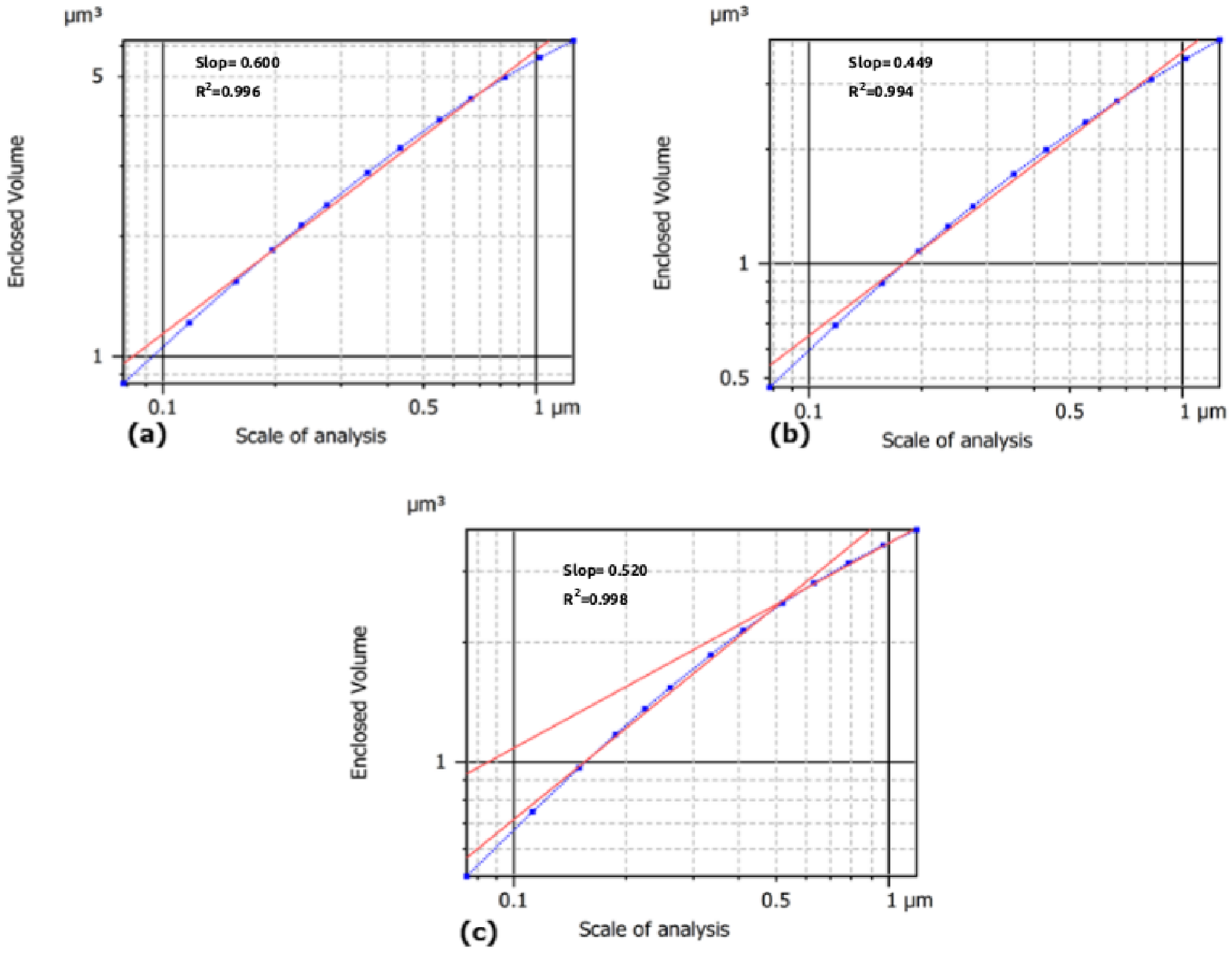} 
 \caption{Enclosed volume for KBr films of: (a) 200 nm, (b) 300 nm and (c) 500 nm. The fractal dimension is obtained from the slope(s) of the graphs.}
     \label{XRD}
    \end{centering}
  \end{figure}

In practice, none of fracture surface profile strictly followed the self-similar properties and self-similarity may be local only. The 
distribution  of surface irregularity altered with the analysed region and in order to get more clear understanding regarding the variation in the shape irregularity appeared on the film's surface, AFM data was also processed by the enclosing box or the morphological envelope method and obtained graphs are shown in Fig. 7(a)-(c). In these graphs, a segmental character of a doubly log–log plot represents that the fracture surface has been characterized by more than one fractal dimension \cite{Chhabra, Dallaeva}. 

The upper and lower envelopes are evaluated from morphological opening and closing using a structuring element which is a horizontal line segment of length $\xi_{env}$. The graph of the calculated volume for surfaces $(V_{\xi_{env}})$ is plotted as a function of the scale (size of the structuring elements): $ln(V_{\xi_{env}})$/$ln(\xi_{env})$. A logarithmic scale is used for the axes, but the values of the graduations are given as dimensional units. Both of these parameters are calculated for two regression lines, one approaching to the points from the left of the graph, the other from the right. The fractal dimension was estimated from the slope of one of the two regression lines that corresponds best (i.e. the one out of the two regression lines whose correlation coefficient is nearer to 1 for a profile and nearer to 2 for a surface). Table 1 summarized the values of fractal dimensions $D_{f}$ , obtained for $5\mu m \times 5 \mu m$ scanning areas of  KBr films.

\begin{table*}[tp]
\setlength{\tabcolsep}{5pt}
\renewcommand{\arraystretch}{1.5}
\caption{The fractal parameters for 5 $\mu$m$\times$ 5 $\mu$m scanning square areas of KBr films deposited on aluminum substrate.}
\begin{flushleft}
 {\begin{tabular}{l*{10}rrr}
\toprule
  No. & Sample& \multicolumn{7}{c}{Correlation function Method} &&& \multicolumn{1}{c}{Envelope method}
\\  
\cmidrule(r){3-9}
\cmidrule(r){10-12}
      & &$w_{_{Th}}(\mu m)$& & $\xi_{_{Th}}(\mu m)$ &        & $\alpha$& & $D_{f_{Th}}$ & && $D_{f_{env}}$ \\
  
\midrule
1.  & 200nm& 0.0694 && 0.6867&& 0.8022&&2.1978 &&&2.29\\
2. &300nm&0.0331&&0.3217&&0.8869&&2.1131&&&2.25\\
3. &500nm&0.0283&&0.3094&&0.8590&&2.1410&&&2.49\\

    \bottomrule
    \bottomrule
 \end{tabular}}
\end{flushleft}
\label{table:nonlin}
\end{table*}

 \section{Conclusion} 
 In this study, we have characterized the surfaces of KBr films by atomic force microscopy and acquired images are utilized for extracting the statistical description and fractal geometry parameters. The morphological analysis, A-F curve and evolution of interface width collectively exhibit that the deposition of more KBr vapour facilitating the crystal growth, increasing the grain size, densifying the film and smoothing surfaces.   

The evolution of surface parameters using a scaling theory shows a strong influence of global roughness and correlation length on the film thickness. However, the computation of short range roughness, correlation length and fractal dimension do not suggest any dependency on film's surface quality and thickness.. Therefore, it may be concluded that the small scale variation in the surface morphology of thermally deposited  KBr film is independent of thickness. The theoretically computed value of fractal dimension using a height-height correlation function is different from morphological envelop method, this discrepancy may be arises due to mathematical stepping. Since in morphological envelop method, a fracture surface is by more than one fractal dimension, while scaling concept assume a self affine nature of film surfaces.

 \section{Acknowledgment}
  This work was partially supported by the Department of Science and Technology (DST) FIST, PURSE and the Indian Space Research Organization (ISRO) under ISRO-SSPS programs. R. Rai acknowledges University Grant Commission (UGC), India for providing financial support.

\end{document}